\documentclass[rnote]{aa} 
\usepackage{graphicx}
\usepackage{txfonts}
\begin{document}
\hyphenation {at-tempt spots rea-ches Ge-min-ga Mi-la-gro km ef-fects sce-na-rio 
mo-del ex-plo-sions reaches value dif-fu-sion res-pect ca-ses vo-lume ge-ne-ra-ted
sca-les pro-tons ex-pe-ri-ments sug-ges-ted co-ef-fi-cient smal-ler rem-nant small
two-fold gi-ven quo-ted still or-tho-go-nal pa-ral-lax hint dra-ma-tic hy-po-the-sis
mo-de-ling con-stant de-duce}

   \title{The local Galactic magnetic field in the direction of Geminga}

   \author{M. Salvati
          }

   \offprints{M. Salvati}

   \institute{INAF--Osservatorio Astrofisico di Arcetri \\
              Largo Enrico Fermi 5, I--50125 Firenze, Italy\\
              \email{salvati@arcetri.astro.it}
             }

   \date{Received / Accepted }

 
  \abstract
   {The Milagro hot spot A, close to the Galactic anticenter direction, 
    has been tentatively attributed to cosmic rays
    from a local reservoir (at a distance $\approx$~100~pc), freely streaming along 
    diverging and smooth magnetic field lines. This is at variance with the 
    geometry of the $\approx$~kpc scale Galactic magnetic field, which is
    known to be aligned with the spiral arms.}
   {We investigate the information available on the geometry of the
    magnetic field on the scales ($\approx$~100~pc) of relevance here.}
   {The magnetic field immediately upstream of the heliosphere has been investigated
    by previous authors by modeling the interaction of this field with the
    solar wind. At larger distances, we use the dispersion measure and the
    rotation measure of nearby pulsars (especially towards the third Galactic
    quadrant). Additional information about the local field towards the North Polar 
    Spur is taken from previous studies of the diffuse radio emission and the polarization
    of starlight.}
   {The asymmetry of the heliosphere with respect to the incoming interstellar
    medium implies a magnetic field almost orthogonal to the local
    spiral arm, in the general direction of hot spot A, but more to the
    south. This is in good agreement with the nearby pulsar data on the
    one side, and the North Polar Spur data on the other.}
   {The local magnetic field on scales of $\approx$~100~parsecs 
    around the Sun seems to be oriented so as to
    provide a direct connection between the Solar system and
    a possible site of the Geminga supernova; the residual angular difference
    and the shape and orientation of the Milagro hot spot can be attributed to
    the field trailing in the wake of the heliosphere.} 


   \authorrunning{M. Salvati}
   \titlerunning{The magnetic field toward Geminga}

   \maketitle
%

\section{Introduction}

The detection by Milagro of anistropies at small angular scales
in the arrival directions of multi--TeV cosmic ray protons (dubbed
hot spots A and B, \cite{milagro}) has stirred a lively debate.
Indeed, there was no surprise in the detection of anisotropies at
the measured level, but the expectation was that such anisotropies
would appear on large angular scales, in agreement with the
diffusion mode which accounts successfully for the propagation
of cosmic rays.
A positive excess in the general direction of hot spot A 
("tail--in" anisotropy) had been already detected by other
experiments [Tibet Air Shower Array (\cite{tibet}) and Super
Kamiokande I (\cite{kamioka})], although the narrowness of
the feature (only a few degrees) had not been noticed before.

Salvati and Sacco (\cite{ss}, hereafter SS) pointed out that
hot spot A is in the general direction of Geminga, and suggested
that a plausible source could be the Geminga supernova remnant 
(SNR) rather than the pulsar. The SNR would be dispersed by
now, and would survive only as an expanding cloud of cosmic rays.
The distance to it could be 
much smaller than the present distance to the pulsar, if
a positive radial velocity is assumed for the latter. Such
a distance could be crossed by diffusion in the time elapsed
since the explosion (at least with crude assumptions about
the diffusion coefficient). Also the energetics 
turned out right, and the energy
dependence of the diffusion coefficient would account for the
hard spectrum ($\Gamma$$\sim$1.45) of the excess cosmic rays.
Drury and Aharonian (\cite{da}, hereafter DA) criticized SS
on the grounds that the assumed diffusion coefficient was
very implausible, and, moreover, a fully diffusive approach
could not account for the narrow angular size of the hot spots.
They suggested instead that some nearby "cosmic ray reservoir"
was connected to the Solar system by a "magnetic funnel":
the cosmic rays could then stream freely along the (diverging 
and smooth) field lines, while at the same time their pitch 
angle distribution would narrow down to the observed value. 
A hybrid scenario was finally proposed by SS:
there the "cosmic ray reservoir" coincides with the
Geminga SNR; the cosmic rays have to diffuse until they reach
the "first useful magnetic line" which drives them to the
funnel and then to the Solar system. The initial diffusion
accounts for the spectral filtering,
the final streaming accounts for the angular distribution.

There is a major caveat, however. The available information
about the geometry of the Galactic magnetic field
(e.g., \cite{manche}) indicates that on scales $\approx$~kpc
the ordered magnetic field is in the direction of the local
spiral arm, and the chaotic component of the field is somewhat
larger than the ordered one. The magnetic funnel scenario,
on the contrary, requires that (on smaller scales 
$\approx$~100~pc) the field is predominantly ordered, and 
directed toward the anticenter. In the following we discuss 
evidence that this {could} indeed be the case.

\section{The local and very local magnetic field}

Information about the magnitude and direction of the magnetic
field immediately upstream of the heliosphere (i.e., in the
very local interstellar medium still unperturbed by the bow
shock) can be gained by modeling the anistropies observed in
several heliopause tracers (see, for instance, Ratkiewicz,
Ben-Jaffel, \& Grygorczuk \cite{jaffel}, and references therein).
One obtains a very local magnetic field of $\sim 1.8~\mu$G, 
oriented within the interval $203^{\circ} < \ell < 231^{\circ}$,
$-58^{\circ} < b < -41^{\circ}$. Note that this analysis is
insensitive to the sign of the field, so that an equally
admissible solution is $ 23^{\circ} < \ell < 51^{\circ}$, 
$+41^{\circ} < b < +58^{\circ}$. The latter solution is
plotted in Fig.~1 as a circle labeled "B near".

\begin{table}
\caption{Nearby pulsars used in the analysis}  
\label{table:1}      
\centering                     
\begin{tabular}{c c c c c c}  
\hline\hline                 
Name & $\ell_{\rm II}$ & $b_{\rm II}$ & Dist & DM & RM \\
     &  degrees &  degrees & pc & cm$^{-3}$~pc & rad~m$^{-2}$ \\
\hline                       
   J2144-3933 & 2.8 & -49.5 & 180 & 3.35 & -2 \\ 
   J2124-3358 & 10.9 & -45.4 & 250 & 4.60 & 1.2 \\
   J0108-1431 & 140.9 & -76.8 & 130 & 2.4 & -0.3 \\
   B0656+14 & 201.1 & 8.3 & 290 & 14.0 & 23.5 \\
   B0950+08 & 228.9 & 43.7 & 260 & 2.96 & -0.66 \\
   J0437-4715 & 253.4 & -42.0 & 160 & 2.64 & 1.5 \\
   B0833-45 & 263.6 & -2.8 & 290 & 68.0 & 31.4 \\
\hline                     
\end{tabular}
\end{table}

In order to explore the field on scales of a few hundreds
of parsecs from the Solar system, we use the dispersion measure
(DM) and the rotation measure (RM) of nearby radio pulsars
(\cite{manche}). We retrieve from the ATNF Pulsar Database 
(\cite{cat}\footnote{http://www.atnf.csiro.au/research/pulsar/psrcat/})
all the pulsars with measured DM and RM, and
distances less than 300 parsecs. There are seven such objects,
listed in order of increasing Galactic longitude in Table~1.
Their distances are obtained either from the annual parallax, 
or (for J0108 and J2144) from the DM and an assumed model of the
electron distribution. 
Given the relatively small volume, we approximate the magnetic field
as a constant vector, fully described by three independent
components, which we find by minimizing the $\chi^2$ between
the observed and the predicted RM\footnote{{Even 
if each pulsar gives only the B component along the
line of sight $\rm \bf (B \cdot n)$, three or more pulsars widely spaced
over the sky are sufficient to constrain B independent of its
direction.}}

$$\rm RM_{\rm pred} = 0.81 DM_{\rm obs} ({\bf B}_{\mu\rm G} \cdot {\bf n})$$

On the other hand, the RM is well known to vary widely
even for small angular displacements, so that the ordered component
of the field is found by averaging the data over large regions of 
the sky. We do not perform any average, given the small number of 
entries, however we must be prepared to find a $\chi^2$ very much higher 
than $\approx$1 per degree of freedom. We use this estimator only to draw
some qualitative guesses. 
The values of DM and RM of two particular objects (B0656+14 and
B0833-45) are by far larger than the other values, as one could
have expected because of their location in the Monogem
and Vela SNR, respectively. This adds one further caveat to our
results, since a dense, young SNR could be dominated by a local
magnetic field of its own.

If we retain in the fit all the seven pulsars, we obtain 
$\rm B \sim 1.9~\mu G$, $\ell \sim 16^{\circ}$, $b \sim 45^{\circ}$
with a reduced $\chi^2$ of around 470~(!). If instead we retain 
only the four pulsars lying in the third Galactic quadrant, since
the excess cosmic rays reach the Solar system from this general direction,
we obtain $\rm B \sim 3.3~\mu G$, $\ell \sim 5^{\circ}$, $b \sim 
42^{\circ}$ with a reduced $\chi^2$ of only (!) 40. 
{As a check on our findings, we have repeated the analysis
for all pulsars with distances less than 500 parsecs: on the
one hand, we become sensitive to the field on scales somewhat
larger than the ones of interest; on the other, we improve the
statistics by increasing the sample to 18 objects in total, and 
to 8 in the third Galactic quadrant. Finally, we fitted all 
pulsars with distances between 500 and 2,000 parsecs (103 objects), 
which should reproduce the azimuthal geometry already established
by previous authors. On such large scales we include a portion
of the Sagittarius -- Carina arm, where the field is known to
reverse direction (\cite{manche}); so, in order to keep the sample
clean, we also fitted a subsample including only the pulsars lying 
outside the arm (57 objects).}

\begin{table}
\caption{Magnetic field obtained from various pulsar samples}  
\label{table:2}      
\centering                     
\begin{tabular}{c c c c c c}  
\hline\hline                 
sample & no. objects & $\ell_{\rm II}$ & $b_{\rm II}$ & B & $\chi^2_{\rm rid}$ \\
 & & degrees &  degrees & $\mu$G & \\
\hline                       
$<300$~pc & 7 & 16 & 45 & 1.9 & 470 \\
III quad & 4 & 5 & 42 & 3.3 & 40 \\
$<500$~pc & 18 & 6 & 28 & 2.7 & 1,500 \\
III quad & 8 & 9 & 43 & 2.5 & 340 \\
0.5--2~kpc & 103 & 80 & -10 & 2.5 & 10,000 \\
arm excl. & 57 & 100 & 6 & 3.1 & 2,200 \\
\hline                     
\end{tabular}
\end{table}

 {The results are summarized in
Table~2. One sees that the $<500$~pc sample gives results in broad 
agreement with the $<300$~pc sample, while the 0.5--2~kpc sample
indicates clearly a rotation of the field which becomes (more
or less) aligned with the Galactic plane in the direction of
the local spiral arm. Note especially that restricting the analysis
to the third Galactic quadrant does not change appreciably the
field, but makes the reduced $\chi^2$ substantially smaller.
The reduced $\chi^2$ becomes substantially smaller also in the large
scale sample, as expected, if one excludes the pulsars inside
the Sagittarius--Carina arm.}

We regard the substantial agreement between the first four sets 
of values in Table~2 as a hint
that our procedure is meaningful. Furthermore, the substantial
agreement between the pulsar derived magnetic field (on scales
$\approx$~100~pc) and the very local, heliopause derived magnetic
field is a hint that in our Galactic neighborhood the
magnetic field is relatively smooth. An independent hint at the 
field smoothness (a prerequisite for the validity of the funnel 
scenario) comes from the very significant decrease of the reduced 
$\chi^2$ in the third quadrant with respect to the all sky value. 
The two $<300$~pc pulsar--derived solutions are plotted in Fig.~1 as two 
crosses labeled "B rm"; we have not computed a confidence region from 
the $\chi^2$ distribution because of the caveats associated with it,
and guess the uncertainty from the difference between the two
solutions.

The structure of the magnetic field towards the Galactic
center is loosely constrained by the pulsar data, which
only suggest a geometry more complex than a uniform field. A 
clearer picture can be obtained by modeling the intensity and
polarization of the nearby extended radio emission (\cite{woll}) 
and the polarization of the light from nearby stars (\cite{frisch}).

The interstellar medium in this general direction has
been perturbed by a series of explosions likely due to
stars in the Sco--Cen association. The radio intensity
and radio polarization maps show the traces of several shells, 
the most prominent of which is the North Polar Spur. One 
of the shells [called "Shell 1" by Wolleben (2007)] may 
have reached the Sun. In order to account for both the
radio and the optical data, the magnetic field in the perturbed 
region is described as a uniform field outside the shells
and, within the shell thickness, as a compressed field lying 
along the meridian circles. The radio data require two
different shells, while the optical data can be fitted with
Shell~1 only, and help to constrain its parameters within
the large radio--derived confidence region.

   \begin{figure}[h!]
   \resizebox{\hsize}{!}{\includegraphics{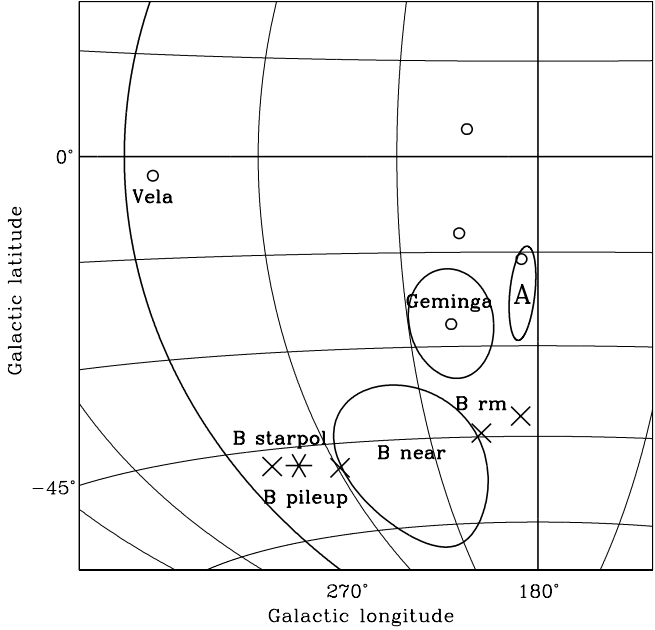}}
   \caption{Aitoff equal area projection in Galactic coordinates
    of the southern half of the third Galactic quadrant. See text
    for the meaning of the symbols. In all cases, the B field is
    directed out of the page towards the reader.}
   \end{figure}

The star symbol labeled "B starpol" in Fig.~1 is the direction
of the uniform field inside which Shell~1 is expanding
(\cite{frisch}, no errors given). This would be the direction 
of the field outside the heliosphere if Shell~1 had not reached
us yet. Otherwise the field would be the one compressed along
the local meridian line of the shell: the two crosses labeled
"B pileup" represent two possible choices of the shell center.
Note the near coincidence of "B starpol" and "B pileup", which
is due to the shell expansion center being at almost 90$^{\circ}$
with respect to "B starpol". 

Figure~1 summarizes our findings. Here the southern half of the
third Galactic quadrant is plotted in an Aitoff equal area 
projection. The various estimates of the B field direction
have already been discussed. {For the sake of comparison,
all of them are represented as the respective points at $-\infty$,
but they pertain to different physical regions: B$_{\rm rm}$
should be valid at $\approx$~100~pc in the third quadrant,
B$_{\rm near}$ and B$_{\rm pileup}$ should be valid only very 
close to the Sun, and B$_{\rm starpol}$ should be valid at
$\approx$~100~pc in the first quadrant. In the latter
region the field has been heavily distorted by the expansion
of the radio shells, however what is plotted here is the unperturbed, 
pre--shell field, so that we can draw meanigful conclusions from its 
smooth connection with B$_{\rm near}$ and B$_{\rm rm}$ (see Section~3
and Figures 2 and 3).}

The hot spot A and the heliotail direction are represented by 
the ellipse labeled "A" and the small dot inside it. Finally, the
three dots in descending sequence are: the present position of
the Geminga pulsar; the position it would have had at explosion
if its motion were parallel to the plane of the sky with the 
measured proper motion value; and the position it would have 
had if the explosion had occurred at the "minimum" distance of 
65~pc [i.e., with a positive 160~km~s$^{-1}$ radial velocity
included, see SS; in both cases, the time 
elapsed since the explosion is assumed equal to the spin down
age of the pulsar, $3.4~10^{5}$~yr (\cite{geminga})]. Around the 
latter dot we have drawn a circle of 10~pc radius, representing 
a fully developed SNR.

\section{Discussion and conclusions}

The first result we want to stress is the geometry displayed in
Fig.~1: the direction of the local magnetic field,
the direction of hot spot A, and the direction to a {possible}
location for the Geminga SNR all lie within a few degrees from
one another\footnote{{Battaner, Castellano \& Masip (\cite{batta}) have 
developed a model for the dipole-like Milagro anisotropy (note: this 
is different from the point-like anisotropy discussed here). Their
model succeeds in accounting for the dipole under the assumption
of a local magnetic field basically aligned with the
local spiral arm, i.e. at a large angle with the one derived
here. But if one assumes a streaming motion of the cosmic
rays along the magnetic field, beside the orthogonal motion 
derived by them based on an ad--hoc turbulent stress, the two
estimates can be reconciled.}}. {Apart from the hot spot, all the 
other directions in Fig.~1 are not directly measured, and are 
obtained by modeling the available datasets, not always plentiful. 
If, nonetheless, we take these results at face value,} one of the 
main objections to a diffusion-plus-funnel scenario could be removed: 
the field on the relevant scales seems to be almost orthogonal to the 
large scale one, and to point in the right direction.

The second result concerns the smoothness of the local field,
which is necessary if the cosmic rays have to stream freely
along the magnetic funnel in order to be focussed within a
narrow range of pitch angles. The evidence for such smoothness
{(admittedly meager)} comes from two findings. One is the
dramatic drop of the reduced $\chi^2$ if one selects for modeling 
only the pulsars lying in the third Galactic quadrant.
The other is the near coincidence 
between the directions of the very local, heliopause derived 
magnetic field ("B near"), and the $\approx$~100~pc scale one, 
either pulsar derived ("B rm"), or radio--optical derived 
(for the unperturbed configuration, "B starpol"). Indeed, 
one notes that there is a regular and smooth "rotation" of the
B field vector: it comes from about the anticenter when the 
field is determined in the third Galactic quadrant; grows in
Galactic longitude by about 30$^{\circ}$ at the Solar system; and 
grows still by another 30$^{\circ}$ when the (unperturbed) field 
is determined in the direction of the Galactic center.

{We sketch the envisaged geometry in Figures 2 and 3. 
They are the projection on the Galactic plane and, respectively, 
the meridian plane $\ell = 180^{\circ}$ of the local and
very local magnetic field, and of Shell~1 of Wolleben (2007). 
The crudeness of the sketch gives the impression of a sharp
bend at the solar position, which would be injustified; but
an equally valid (and equally arbitrary) representation could
involve magnetic lines with a curvature radius as large as the 
Figures themselves. Also, the actual rotation in three
dimensions amounts to 46 degrees only; this is strongly
amplified by projection effects. Finally, the dashed lines
inside the shell refer to the pre--shell situation: after
the shell has overtaken them, they are draped along the
shell surface.}

   \begin{figure}[h!]
   \resizebox{\hsize}{!}{\includegraphics{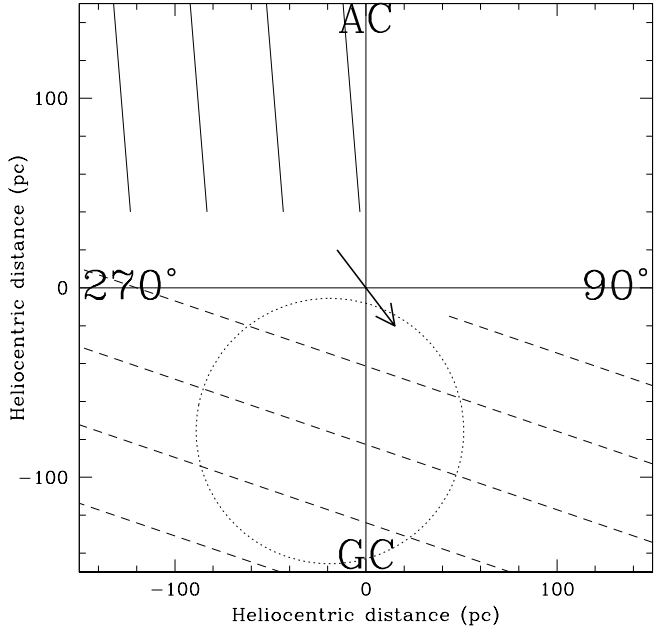}}
   \caption{Orthogonal projection on the Galactic plane of
    the pulsar derived field (the rightmost cross of Fig.~1,
    solid lines), the unperturbed radio--optical derived field
    (the star symbol of Fig.~1, dashed lines), and the radio
    Shell~1 (under the assumption that it has not reached
    the Sun yet, dotted circle). The heavy line through the center
    is the heliospheric derived, very local field. Its arrow
    indicates the field orientation. The axes are labeled GC
    (Galactic center), AC (anticenter), $90^{\circ}$ and 
    $270^{\circ}$ (for the Galactic longitude).}
   \end{figure}

   \begin{figure}[h!]
   \resizebox{\hsize}{!}{\includegraphics{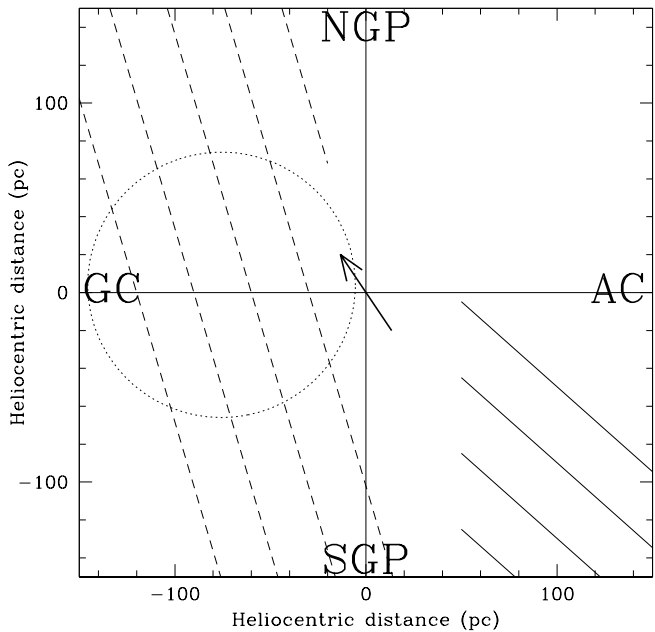}}
   \caption{The same as Fig.~2, for the meridian plane 
   $\ell = 180^{\circ}$. The vertical axis is labeled NGP
   (North Galactic pole) and SGP (South Galactic pole).}
   \end{figure}

We do not regard as a major discrepancy the residual angular
separation between the {assumed} direction of the Geminga SNR, 
"B near", "B rm", and the actual position of hot spot A. However, 
some plausibility arguments can be given which could account for 
the discrepancy.

As argued in SS, the SNR responsible for the "cosmic ray 
reservoir" (Geminga or other) should be close to the magnetic 
funnel, so that diffusion with reasonable coefficients could 
account for the propagation of the cosmic rays from
the SNR to the funnel in the time elapsed since the explosion. 
At the same time, however, the SNR should not lie directly on
the "first useful field line", otherwise one would miss the
energy filtering (needed to explain the spectral hardness of 
the cosmic ray excess).

Second, a small meandering of the magnetic field, sufficient to
account for the angular difference between "B rm" and "B near",
is not only plausible, but indeed very likely. The important
point is that such small deflections over several tens of
parsecs are by far insufficient to affect the free streaming
of the cosmic rays.

Third, the actual position of hot spot A is perhaps determined
by the direction of the very local magnetic field in the wake
of the heliosphere. Indeed, the direction of "B near" depicted in
Fig.~1 refers to the field ahead of the heliosphere, before any
interaction with it (Ratkiewicz et al. \cite{jaffel}). After the 
wind, the field should 
become more aligned with the heliotail, and it is plausible than
the alignment lasts for several times the distance to the heliopause,
i.e. for about $\lesssim~10^{16}$~cm. Such a distance is comparable
with the Larmor radius of a 10~TeV particle; the radius of curvature
needed for a 20$^{\circ}$ swing over this distance is of course
larger still, so that the free streaming of the cosmic rays should
not be disrupted.

Note that the ambient magnetic field lines will tend to wrap around
the heliosphere in the plane passing through the apex and containing
the field and wind directions, while they will 
tend to slip apart on the two sides; the cosmic rays 
will then be focused in the said plane, and de-focused on the two 
sides. This corresponds {roughly} to the elliptical shape and
the position angle of hot spot A. Qualitatively, the pile up of
the lines toward the heliotail could also account for the gradient
observed in hot spot A along the major axis, with the maximum
on the heliotail side.

{The geometry of the magnetic field which we have discussed
thus far is perhaps too detailed in comparison with the available
evidence. Still, it is by far insufficient for a quantitative 
estimate of the anisotropy amplitude. In order to achieve this,
one should follow with high spatial and temporal resolution the
expansion of the cosmic ray cloud injected by the Supernova, 
including the individual field irregularities throughout the
cloud volume. The cloud, which we assume spherical, could well
be elongated in one dimension, or have a complicated topology.
The best we can do at the moment is to show that 
the observed anisotropy can be sustained by a minuscule gradient 
in the density of the cosmic rays, a gradient not implausible 
for a location relatively close to a relatively recent Supernova. 

We write the energy flux measured from hot spot A (\cite{milagro}) 
as follows

\begin{equation}
\Phi\sim 5~10^{-4} \times 6.7~10^{-6} \sim 3.3~10^{-9}~
\rm erg~cm^{-2}~s^{-1}~sr^{-1}
\end{equation}

The magnetic funnel at the injection side is about 20 times
narrower than at the Sun side (see DA), and the particle pitch
angle squared scales inversely by the same factor, so that $\Phi$
is constant. Hence the required density is

\begin{equation}
\rm n \sim 4\pi \frac{\Phi}{\rm c} \sim 1.4~10^{-18}~\rm erg~cm^{-3}
\end{equation}

If the Supernova explosion injects $10^{50}$~erg in cosmic rays
with the same spectrum as the general cosmic ray population, the
10--TeV reservoir amounts to 1.7~$10^{47}$~erg. Spreading this
reservoir in a sphere of radius 100~pc (the length of the
funnel suggested by DA, and a plausible distance for the
Geminga explosion) one gets n~$\sim 1.4~10^{-15}~\rm erg~cm^{-3}$,
i.e. three orders of magnitude larger than Eq.~(2).

Conversely, we can compute the cloud volume corresponding to the
density of Eq.~(2), $\rm V\sim1.210^{65}\rm cm^{3}$, and 
deduce a diffusion coefficient. Setting the time t since the
explosion of Geminga equal to $3.4~10^{5}$~yr (Bignami \& Caraveo
1996), we find

\begin{equation}
\rm D = \big(\frac{3\rm V}{4\pi}\big)^{2/3}\times \frac{1}{4\rm t}
\sim 2.2~10^{29}~\rm cm^2~s^{-1}
\end{equation}

The above value for D is not far from what is usually assumed 
in cosmic rays modeling (e.g. \cite{blasi}), and is another 
plausibility argument in favor of our suggestion: hot spot A 
could be the first example of direct cosmic ray astronomy.}

{}

\end{document}